\documentclass{article}

\usepackage{arxiv}

\usepackage[utf8]{inputenc} 
\usepackage[T1]{fontenc}    
\usepackage{hyperref}       
\usepackage{url}            
\usepackage{booktabs}       
\usepackage{amsfonts}       
\usepackage{nicefrac}       
\usepackage{microtype}      

\usepackage{cite}
\usepackage{amsmath,amssymb,amsfonts}
\usepackage{graphicx}
\usepackage{enumitem}
\usepackage[usestackEOL]{stackengine}
\usepackage[dvipsnames]{xcolor}

\title{Andro-Simnet: Android Malware Family Classification using Social Network Analysis}

\author{
  Hye Min Kim  \\
  Graduate School of Information Security\\
  Korea University\\
  Seoul, Republic of Korea \\
  \texttt{hm941224@korea.ac.kr} \\
  \And
  Hyun Min Song \\
  Graduate School of Information Security\\
  Korea University\\
  Seoul, Republic of Korea \\
  \texttt{signos@korea.ac.kr} \\
  \And
  Jae Woo Seo\\
  Software R\&D Center\\
  Samsung Electronics\\
  Seoul, Republic of Korea \\
  \texttt{jaewoo13.seo@samsung.com} \\ 
  \And
  Huy Kang Kim \\
  Graduate School of Information Security\\
  Korea University\\
  Seoul, Republic of Korea \\
  \texttt{cenda@korea.ac.kr} \\  
}

\begin{document}
\maketitle

\begin{abstract}
While the rapid adaptation of mobile devices changes our daily life more conveniently, the threat derived from malware is also increased. There are lots of research to detect malware to protect mobile devices, but most of them adopt only signature-based malware detection method that can be easily bypassed by polymorphic and metamorphic malware. To detect malware and its variants, it is essential to adopt behavior-based detection for efficient malware classification. This paper presents a system that classifies malware by using common behavioral characteristics along with malware families. We measure the similarity between malware families with carefully chosen features commonly appeared in the same family. With the proposed similarity measure, we can classify malware by malware’s attack behavior pattern and tactical characteristics. Also, we apply a community detection algorithm to increase the modularity within each malware family network aggregation. To maintain high classification accuracy, we propose a process to derive the optimal weights of the selected features in the proposed similarity measure. During this process, we find out which features are significant for representing the similarity between malware samples. Finally, we provide an intuitive graph visualization of malware samples which is helpful to understand the distribution and likeness of the malware networks. In the experiment, the proposed system achieved 97\% accuracy for malware classification and 95\% accuracy for prediction by K-fold cross-validation using the real malware dataset.
\end{abstract}


\section{Introduction}

Recently, the number of malware is increasing day by day, and it is generating many victims. Since a growth rate and a spreading rate of malware are growing, a manual malware analysis method has become hard to cope with it. Therefore, a process of analysis, classification, and detection of malware need to be automated \cite{ahmadi2016novel}. Furthermore, as an evasion and obfuscation technology advances, malware becomes more sophisticated and complex and then easily bypasses signature-based anti-virus programs. A typical example of passing through them is polymorphic malware. It autonomously modifies the original malware and generates variants to avoid signature-based detection measure. To solve this problem, we need to use behavioral characteristics of malware to detect malware which has the same attack patterns. Behavior-based detection can easily reveal the given malware’s variants because the malware samples in the same family are born to achieve the same attack goal and tactics. It means malware in the same family is likely to have similar function call sequences. There have been researches that attempt to detect malware based on the similarity, and they use only one or few simple features such as permission \cite{schmidt2009static} and call graph \cite{zhang2014semantics}. Because it is difficult to find out an association between features, it is not well fit to modern malware clustering. Especially, permission information is a no longer useful feature because even benign apps require lots of permissions. Even worse, if detection systems only employ one or small type of features, then it yields biased results because it observes only one aspect of malware. To inspect whole aspects of malware, a hybrid analysis system that combines various features from the signature-based analysis and behavior analysis system is required.

We measure and compare the similarity with the various features extracted from the static and dynamic analysis. All extracted data are vectorized to quantify the malware’s behavioral similarity as a numeric value for the comparison. We design a vector that can well describe the behavioral pattern of malware. Among many features, we use API call sequences which can represent malware’s behavioral patterns. Then, we normalize the string of API call sequences because the length of the sequence could vary by malware. We create a new feature that is the value quantified as a similar relation between malware samples, using whole raw data without missing information. We handle all of the similarity values between all of the selected features for a machine learning process. The final similarity between malware samples is calculated as the weighted sum of these proposed feature similarities. A weighted sum of all feature’s similarity can express overall similarity relation between malware samples easily, but still needs adjustment of the weight values.

In general, the weight values of multiple similarity factors are initially given by human experts. In many research, they are usually set with the evenly divided value or heuristically given value whether the features of vectors are orthogonal or not. These values need to be updated when new malware sample is incoming to classification systems. However, in many research, the process for gaining optimal weight value is omitted. To maintain the classification system's performance high, we apply a community detection algorithm of the social network analysis method to derive the optimal weights. 

Classifying malware into a specific family implies that each malware sample in the same family has similar characteristics between the classified samples. Since we perform clustering in a direction increasing the correlation between similar malware samples, the community detection algorithm is an appropriate method. This algorithm makes the modularity of each network aggregation the highest value.

In this paper, we propose the Andro-Simnet that classifies Android malware family with malware similarity network graph. Also, we can conclude which features are essential for classifying malware with the optimal weights. In this work, we use malware samples on Android platform provided by Korea Internet and Security Agency (KISA) and the dataset is available at the website \cite{dataset}.

The key contributions of this paper are as follows:

\begin{itemize}
	\item We use the hybrid system that uses both of the results from the static analysis and dynamic analysis. Thus, we can use signature-based information and behavior-based information to classify malware, while most other papers use only one or two features with limitations. It will be helpful to detect and response to the variants of malware. 
	\item In addition to the extracted features from the static and dynamic analysis, we used the similarity value derived from the social network graph features. To represent relative weight value between the selected features, we use the community detection algorithm. In our approach, we can cluster malware more precisely. 
	\item We drive the optimal weight values by the proposed machine learning algorithm, not by a heuristic decision as other research. Thus, our malware classification system can continuously update the weight value between selected features optimally and automatically. Also, without human expert knowledge, it makes easy for analysts to choose optimal weight values for features during the classification process.
	\item Finally, we test our system with the real-world malware samples and predict one’s family in using K-fold cross-validation. Through these results, we validate that Andro-Simnet can easily classify and make groups for malware samples. It can help to find the newly discovered malware’s attack goal by checking with the most similar known malware family’s purpose. Finally, using the Andro-Simnet, it is possible to react to the variants of malware quickly and accurately.
\end{itemize}

\section{Related work}

We use Android malware samples to find the similarity of malware and classify it into families. We investigate the method of measuring similar-relation of Android malware and proceed literature study about the system to classify or detect malware. With similar studies, we can consider what features are used to measure similarity and how to derive the final similarity.

Jang et al. categorized Android malware by measuring three similarities. First, they selected the specific APIs that is frequently used by known malware and checked if there is the sequence of the APIs that is related to the specific API. And they measured the similarity using the Needleman-Wunsch algorithm for API sequence converted into ASCII code format. Next, the similarity of the used commands was calculated by applying the Jaccard distance algorithm. Finally, the similarity of the required permission was calculated by applying the Levenshtein distance algorithm by reference of whether important permission is requested. Then, the malware was classified according to the largest similarity among three similarities. The blacklist is updated whenever a new malware sample is received \cite{jang2016andro}. The best results were obtained in the malware detection system using both static analysis features and dynamic analysis features. As this paper is our key paper, we refer to the feature of malware and the similarity measure from it. Therefore, we can use hybrid features to classify malware.

Ding et al. explained that Android API features and required permission are not powerful enough to counter malware’s obfuscation techniques because they cannot represent dependencies between two features. In this study, they extracted the structural feature of the Android application from the function call graph. They also proposed the classification system using that feature. The classification system of this study used required permissions, function calls in the main classes, and function call structures as features. The classification system developed by Ding shows higher accuracy than the classification system using only required permission and Android API calls \cite{ding2016android}. It can be seen that the structural features are used to explain the similarity of malware, even when dynamic analysis results are excluded. They decided malicious API feature heuristically, that means the expert is needed to use their system. Compared with our study, our system derives the optimal weights of each similarity feature automatically so that the knowledge of the expert is not necessary.

Zhong et al. proposed a malware classification method considering the metamorphism of malware programs to find the most similar family. The author used three numeric features APIs, referenced strings, and basic blocks and integrated each of the similarities as the arithmetic average \cite{zhong2012malware}. This methodology is similar to our system in that it can integrate various features, but it can be deteriorated because of non-weighted similarity. Conversely, we use feature-based-weighted similarity derived automatically from the Andro-Simnet.

DroidScribe used runtime behavior derived from system calls observed during dynamic analysis to classify Android malware into families. It used the hybrid prediction methodology: CP augmenting SVM, which achieved better classification accuracy. It showed that dynamic analysis is resilient to obfuscation and use of native code and can be a useful feature for Android malware family classification \cite{dash2016droidscribe}. Therefore, we can get a sight to use dynamic features to classify malware samples.

Pehlivan et al. used classification algorithms in permission-based Android malware detection. They made features with permission characteristics, version code, and version name. The feature set of an application can be specified as a feature vector which includes 182 features. They used four feature selection algorithms and five classification algorithms for evaluating their classification. As a result, it gave a better performance to use 25 features and increasing the number of features to 50 did not improve the performance. They also found that the tree-based algorithm has better performance than the Bayesian algorithm \cite{pehlivan2014analysis}. They used more than 25 features to classify malware, but we use only four features for the same purpose. Fewer features can make time-efficiency higher, and features that we use is six times less than theirs.

Faruki et al. measured the statistical similarity of code blocks of zero-day samples using a syntactic foot-printing mechanism to detect Android malware. They had proposed a similarity to use digest hashing mechanism for detecting code obfuscated malware. In the train set, the malware was detected with the accuracy of 98\% to 99\%, however, in the test set, the accuracy of detection was 72\%, and the code obfuscated malware was 63\% training \cite{faruki2013androsimilar}. Thus, there is a limit to detect malware with only static analysis result, and a model adjusted to the training set generated the biased result. Their system is hard to classify new malware samples because of the biased features. We conduct the K-fold classification to examine how well it can classify new malware samples. After this experiment, we can get higher accuracy. 
\section{Methodologies}

\begin{figure}
	\centering
	\centerline{\includegraphics[width=0.8\textwidth]{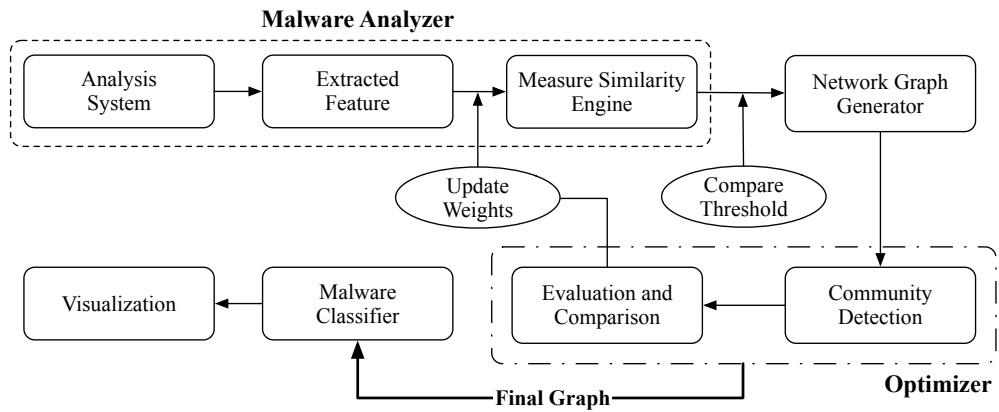}}
	\caption{Overall process of Andro-Simnet}
	\label{fig:fig1}
	\vspace{-0.2cm}
\end{figure}

\begin{figure}
	\centering
	\centerline{\includegraphics[width=0.8\textwidth]{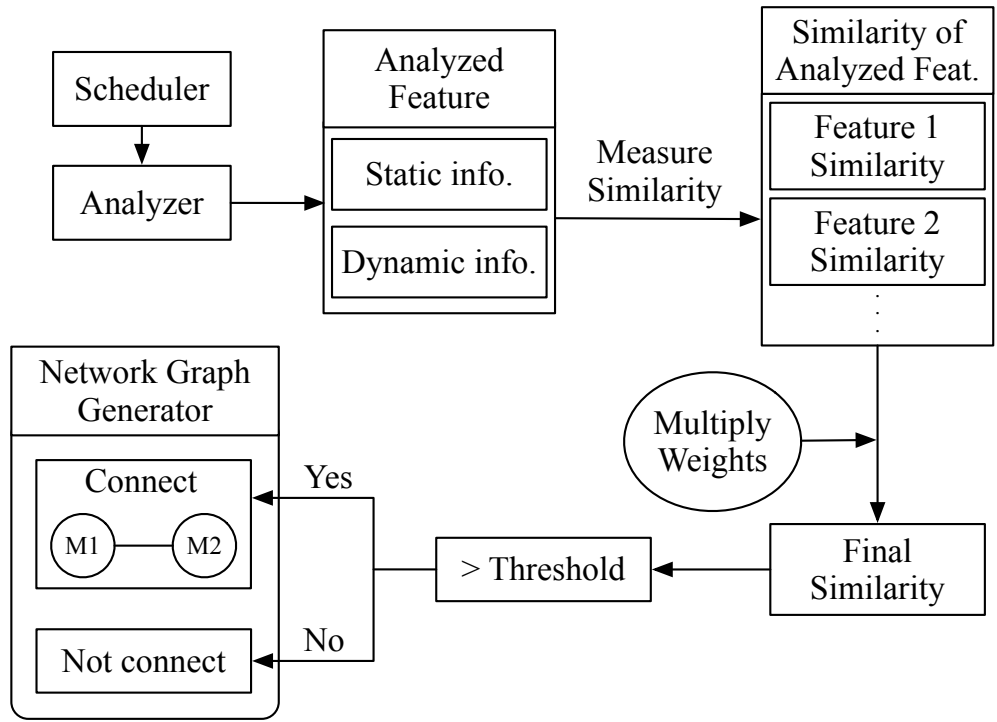}}
	\caption{The process of Analyzer and Network Graph Generator}
	\label{fig:fig2}
	\vspace{-0.2cm}
\end{figure}

The process overview of the Andro-Simnet is described in Fig.~\ref{fig:fig1}. Analyzer extracts static features and dynamic features of malware. The similarity of each feature is measured by different methods depending on the characteristics of the features. Each feature similarity is multiplied by a weight and summed up to calculate the final similarity. Network Graph Generator makes node-edge relations for similar malware samples with this. The node in the network graph represents each malware sample. Nodes that the final similarity between them is higher than the given threshold is connected with the weighted edge. It means that connected malware nodes are similar as much as the value of the final similarity. After the network graph is built, community detection algorithm clusters malware samples into the families. The clusters are stored separately for evaluation during clustering. The process of Optimizer is repeatedly performed to adjust weights until the optimal weights are derived. When the weight is no longer adjusted, the repetition is stopped. The final graph is built based on the optimal weights. Through Malware Classifier, malware nodes are classified into their families. After these processes are ended, we can easily understand classification result with force-directed network graph colored as clusters. The details of each step are described in the following sections.

\subsection{Feature Selection and Measure the Similarity of Malware}

We implement an automated hybrid-analysis system that is based on Cuckoo system to extract static features and dynamic features from malware \cite{CuckooDroid}. The Scheduler manages malware analysis tasks by creating a task queue and assigning the task to the idle virtual machine. The Analyzer performs static analysis and dynamic analysis about 10 minutes. Then, we can get the API call sequence, network information, Android method and so on. We measure the similarity of each feature that is selected among the analyzed results. The process as mentioned above is shown in Fig.~\ref{fig:fig2}.

\textbf{Feature selection} There are various features extracted from the analysis system (e.g., network packet, strings that used in malware, a method call). When we select four features, we consider that both malware behaviors and the signatures of the attacker are useful to represent the similar-relation of malware. The chosen features and the details of the reason why we select that feature are described below:
\begin{itemize}
	\item \textbf{API call sequence} As we mentioned in the introduction section, the API call sequence is closely related to the attack goal of malware. The API call sequence cannot be extracted by code analysis and package analysis. We can get it only by executing malware in the operating system. Through dynamic analysis, it is possible to grasp the behavior of malware in order of time to achieve its attack goal. API call sequence is the most important feature because it can figure out the malware's functional information. We can infer the malware’s function to achieve its attack goal by using this information because malware behavior is strongly related to the attack goal. Therefore, if the similarity of API call sequence is obtained, malware samples with similar functions can be classified with higher accuracy.
	\item \textbf{Permission} Because Android applications operate in a single process sandbox, they need to declare the necessary permissions that are not provided by the default sandbox. The permission is declared according to the attack goal of the malware (e.g., when the attacker’s goal is to get user information from the smartphone, the permission that can access data in the system is declared). It causes malware to gain system-level privileges or access authority. Thus, the higher the number of the permissions declared, the more dangerous malware is. Therefore, the permission can be the feature to classify similar families according to the purposeful attack functions.
	\item \textbf{File name} There are image files, XML files, source code files and any others in the Android package. Because there are specific rules when naming the file according to the developer's propensity, the name of the file includes the signature information of the developer. The same signature information may remain in the application package while the developer is creating other applications. Therefore, the name of the file can be used for comparing the package of malware and obtaining the similarity of it.
	\item \textbf{Activity name} The name of the activity is an application component, and the interface window is given for each activity. Usually, several activities work organically and operate mainly on the main activity. The name of the activity, such as the name of the file, is decided by the developer. Since the activity also includes the package name of the application, the developer's signature information in the activity name is more inclined than that in the file name. Also, as the activity name is referred from outwards, it cannot be changed when polymorphic malware transforms its code automatically. Therefore, to measure the similarity of malware samples, activity names are selected as the signature-based feature.
\end{itemize}

\textbf{Measure the similarity of malware} The similarities of selected features are measured by different methods depending on the characteristics of each feature. The methods to measure two types of data are described below:
\textbf{a)Sequence} An algorithm used to measure sequence similarity should satisfy the following conditions.
\begin{itemize}
	\item The length of the sequence should not significantly affect the similarity.
	\item It should be possible to have better performance in measuring the similarity between long sequences.
	\item The similarity should not vary significantly due to slight differences in content. 
\end{itemize}

Considering these requirements, we adopt the Locality Sensitive Hashing (LSH) algorithm. The LSH algorithm is designed to maximize the collision probability for similar inputs in the process of making the input of different lengths the output which has equal length. The conception of the LSH algorithm is different from cryptographic hashes, that has an entirely opposite purpose. We select the Nilsimsa hash algorithm, which is one of the LSH algorithms, to measure the similarity of API call sequence \cite{damiani2004open}. At first, API call sequences of malware are encoded to digests of the same length. Then, the similarity is measured by comparing those digests of 256 bits and normalized at the value between -1 and 1. 

\textbf{b)Set of strings} The remaining three features consist of a set of strings. The Jaccard similarity is widely used to compare the similarity and diversity of given sets. The basic concept of the Jaccard similarity is defined by dividing the size of intersection by the size of the union of two sets A and B, and the equation is as follows:

\begin{equation}
Jaccard Similarity(A,B) = \frac{|A \cap B|}{|A \cup B|}
\end{equation}

Jaccard similarity is effective regarding performance and has a value between 0 and 1. The value of 0 means that the given two sets have no common element and the value of 1 means that the two sets are equivalent. 

\textbf{Calculate the final similarity} The final similarity of malware is calculated as the weighted sum of feature similarities, and the equation is as follows:
\begin{equation}
FS = \sum_{i=0}^{n} W_{i} S_{i}
\end{equation}

Where $s_{i}$ is the similarity of each selected feature, $w_{i}$ is the weight of each similarity, and $FS$ is the final similarity. For finding the optimal weights, the next section explains the process of machine learning method.

\subsection{Clustering Malware by Social Network Analysis}

\begin{figure}
	\centering
	\centerline{\includegraphics[width=0.8\textwidth]{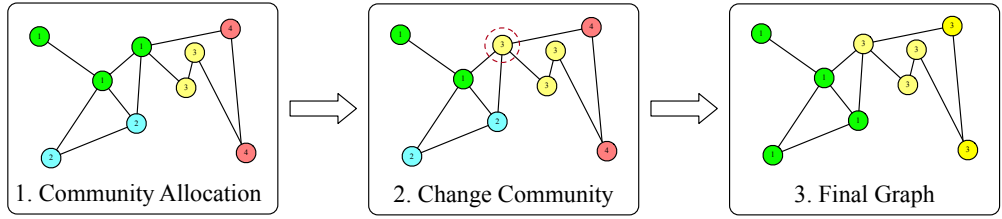}}
	\caption{Louvain community detection algorithm}
	\label{fig:fig3}
	\vspace{-0.2cm}
\end{figure}

\textbf{Social network analysis} In this paper, the feature used for clustering represents a value indicating the similarity between malware samples, not the characteristics of specific malware one by one. Thus, it is difficult to classify malware by applying existing machine learning algorithms to clustering. We focus on how similar malware samples included in the same family are. Considering this condition, it is necessary to use a social network analysis method to conjugate the connection relationship of malware nodes. We obtain the similarity of malware and then multiply weights to the similarities for obtaining the final similarity of malware. When the final similarity exceeds the threshold, the edges are connected to create a network graph. With the given network graph, the social network analysis method performs the malware clustering. We use the Louvain community detection algorithm in this paper that detects communities by iteratively optimizing local communities until global modularity no longer increases. Because each clustered community is assigned to the class of the largest number of nodes in the community, we need to consider a large number of malware samples for scalability. The Louvain community detection algorithm described in Fig.~\ref{fig:fig3} outperforms other algorithms computationally intensive in large networks \cite{blondel2008fast}. 

In Community Allocation phase, each node is assigned to different communities according to the connection of nodes. In the Change Community phase, the community of node \textit{i} into the community is changed to which community node j belongs. The modularity of the graphs that are before and after the allocation is compared. If the modularity of the current graph is more than that of the previous graph, the current graph is sustained. In contrast, the graph is returned to the former. These whole processes of optimization are repeated over and over until getting the highest degree of modularity. The community relationship of the final model would be defined as these processes no longer increase the modularity. A high degree of modularity means that the association of nodes in the community is strong. When the modularity of each community is the highest, it creates a final network graph classified by each community.

\textbf{Clustering malware with the optimal weights}: To derive the optimal weights, Optimizer performs the following processes. When the community detection algorithm clusters a given network graph, the $Error$ of clustering result is calculated according to the ground truth. When the error of current clustering $Error_{current}$ is smaller than the error of the previous clustering $Error_{previous}$, the weights $W_{current}$ at the current iteration is stored as the optimal weights $W_{optimal}$. For the next step of optimization, we update weights as much as the learning rate. 

\begin{figure*}
	\centering
	\centerline{\includegraphics[width=0.8\textwidth]{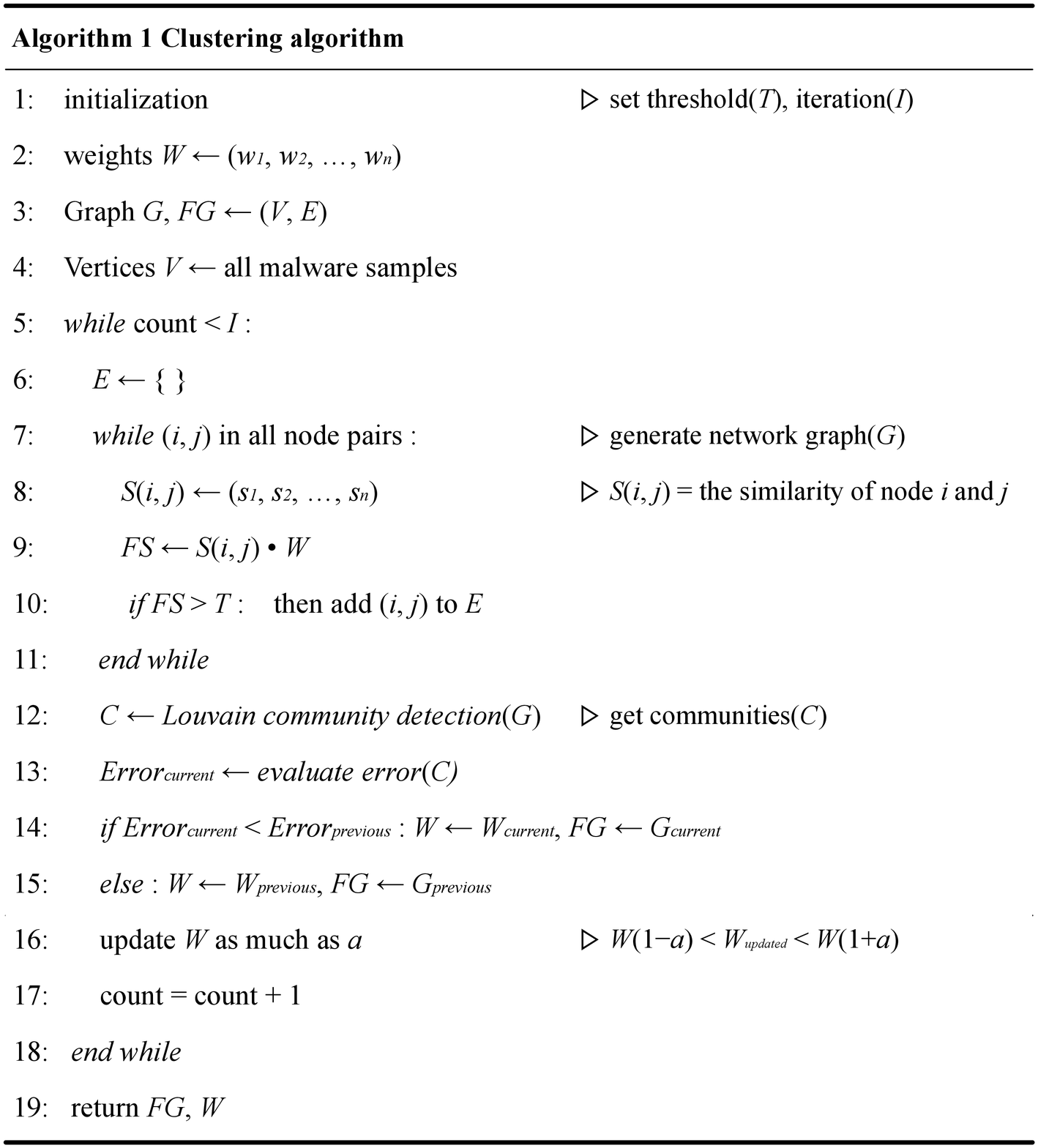}}
	\caption{Louvain community detection algorithm}
	\label{fig:fig4}
	\vspace{-0.2cm}
\end{figure*}

Finally, we define the optimal weights after finishing the final iteration and then generate network graph and use it for deriving the final accuracy of the clustering. Fig. \ref{fig:fig4} shows the pseudocode of the algorithm for clustering malware and it is repeated as iteration $I$.

\subsection{Visualization of Malware Classification}

We use the force-directed graph to visualize classification result \cite{bostock2011d3}. The force-directed graph represents the attracting force between nodes with the weighted edge. The higher the weight of the edge between nodes is, the closer the two nodes are placed. According to the purpose of expressing the similarity of malware, rather than only showing the set of connected nodes, the graph should indicate that nodes which are classified as the same family should be located closely. Since the adjacent nodes in the graph mean that they are highly related, it is helpful for users to understand the similarity between nodes intuitively.
\section{Experiment}

\subsection{Implementation}

The Analyzer is based on the Cuckoo analysis system that is open source. The data that cannot be extracted through Cuckoo analysis system must be obtained by an additional method (e.g., we use ‘aapt’ command which allows reading resources and build-environment of Android so that we can get information of required permission). The virtual emulator implemented for dynamic analysis uses the Android 4.1.2 (API 16) version of the SDK platform and has ARM EABI v7a CPU. We perform Analyzer on a remote server. The CPU specification of the remote server is Intel Xeon CPU E5-1650 3.60GHz, and the RAM specification is 32GB. There are two virtual machines on the server with 64bit Ubuntu 16.04 operating system, one is used as a database for experiments, and one is used as the Analyzer. The codes for clustering through social network analysis and evaluating clustering results are implemented in python using python-igraph module.

\subsection{Experiment Setup}

Reference labels of malware samples are diagnosed by BitDefender as the ground truth. It is referred to obtain an accuracy of classification. After the pre-processing of malware samples, they consist of 8 families, and the total number of samples is 682. We exclude samples that are undiagnosed and do not have one or more of features needed to get the similarity. Table.~\ref{tbl:Table1} shows the number of samples according to the label.

\begin{table}
	\caption{REFERENCE LABEL}
	\centering
	\begin{tabular}{|c|c|c|c|}
		\hline
		\textbf{Label} & \textbf{\# of Samples} & \textbf{Label} & \textbf{\# of Samples} \\ \hline
		FakeBank & 192 & SmsSpy & 118 \\ \hline
		Gepew & 114 & Bankun & 91 \\ \hline
		Gidix & 49 & Misosms & 48 \\ \hline
		FakeInst & 46 & Telman & 24 \\ \hline
	\end{tabular}
	\label{tbl:Table1}
\end{table}

\subsection{Process of Experiment}

We insert them into the task queue of the Analyzer through the Scheduler. The Analyzer performs static and dynamic analysis for each sample for 10 minutes. All analysis results are stored in the database so that they can be retrieved when necessary. While the Analyzer is running, the similarities of each feature between malware samples are calculated and stored in the database. We select four features such as API call sequence, the activity name, the file name, and the required permission. The similarity of the API call sequence is obtained from the Nilsimsa hash comparison. In the case of the set of strings, which are the activity name and the file name and the required permission, are obtained using the Jaccard similarity. We multiply the weights to the degree of similarity for each feature, and then the final similarity is calculated. We set the total iteration of Optimizer to 10,000 for the classification with all malware samples and 1,000 for the prediction using K-fold cross-validation. We measure the accuracy varies according to the threshold t (80\% - 95\%) for getting better accuracy in two situations. After the experiment, the optimal threshold is set. The final graph and the optimal weights are derived at the end.

\subsection{Experiment Results and Discussion}

We evaluate the accuracy of the Andro-Simnet in the classification and the prediction of malware samples. Also, we propose how to cope with the limitations. The details of the experiment are described below:

\textbf{Classification}

First, the classification experiments are repeated to determine the optimal threshold. The threshold is increased as 1 from 80 to 95. The iteration of the optimizer is set by 10,000 and initial weights are set by 0.25 equally. 

\begin{figure}
	\centering
	\centerline{\includegraphics[width=0.5\textwidth]{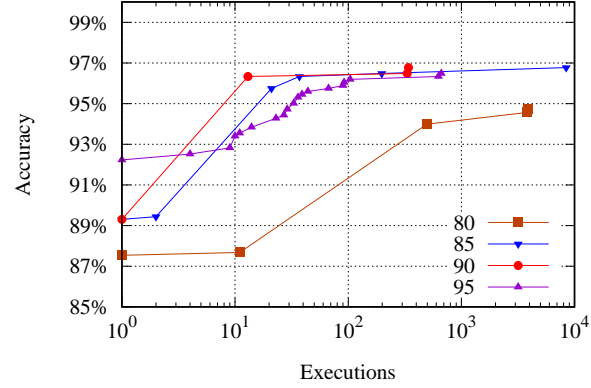}}
	\caption{The change of accuracy according to executions by threshold}
	\label{fig:fig5}
	\vspace{-0.2cm}
\end{figure}

\begin{figure}
	\centering
	\centerline{\includegraphics[width=0.5\textwidth]{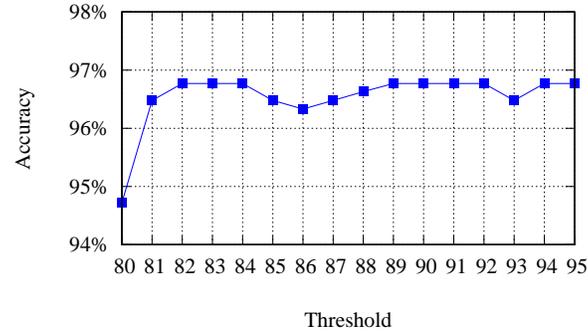}}
	\caption{The final accuracy of the experiment according to threshold}
	\label{fig:fig6}
	\vspace{-0.2cm}
\end{figure}

In each experiment, we can see that weights are fitting in a direction to reduce the errors. Fig.~\ref{fig:fig5} shows the increasing accuracy of the experiment according to each threshold. The point on this graph is the spot that updates the optimal weights as those of the current iteration. The updating point tends to increase as the threshold value increases. The connection between nodes is broken and relatively weak when the threshold is increased. Therefore, nodes that are united tend to be spread. It causes to increase the probability that a node belongs to another community. As the number of community reassignments of nodes increases, the error value changes significantly according to this change. Conversely, if the threshold is getting smaller, the edge between nodes increases. Thus, a strong connection is established with nodes in the community. This means that the node has a lower chance to change its belonging community to others. If there is a change in the community, errors tend to change with a large gap because many nodes are disconnected at the same time.
The final accuracy is shown in Fig.~\ref{fig:fig6}. The results show that the final accuracy is only slightly different according to the threshold. Thus, it is difficult to set the optimal threshold by only referring to this experiment. Therefore, we divide the samples into a training set and a test set and proceed through K-fold cross-validation. 

\textbf{Prediction with K-fold Cross-validation}

In this experiment, we design the model with a high prediction accuracy for a new malware sample. For this purpose, the number of samples is divided into a training set and a test set. The machine is fitting the model with the training set and then the accuracy is measured using the test set. Since the malware samples used in this paper have small numbers to be divided into two sets, we conduct this experiment through K-fold cross-validation that complements the shortage of samples.

\begin{table}
	\caption{THE RESULT OF K-FOLD CROSS-VALIDATION}
	\centering
	\begin{tabular}{|c|c|c|}
		\hline
		\textbf{Iteration(K)} & \begin{tabular}{@{}c@{}}\textbf{Accuracy of} \\ \textbf{classification}\end{tabular}& \textbf{Accuracy of} \\ \textbf{prediction} \\ \hline
		$1^{st}$ & 95.60\% & 95.59\% \\ \hline
		$2^{nd}$ & 97.25\% & 94.85\% \\ \hline
		$3^{rd}$ & 95.79\% & 93.38\% \\ \hline
		$4^{th}$ & 96.70\% & 97.06\% \\ \hline
		$5^{th}$ & 96.88\% & 96.38\% \\ \hline
	\end{tabular}
	\label{tbl:Table2}
\end{table}

\begin{figure}
	\centering
	\centerline{\includegraphics[width=0.5\textwidth]{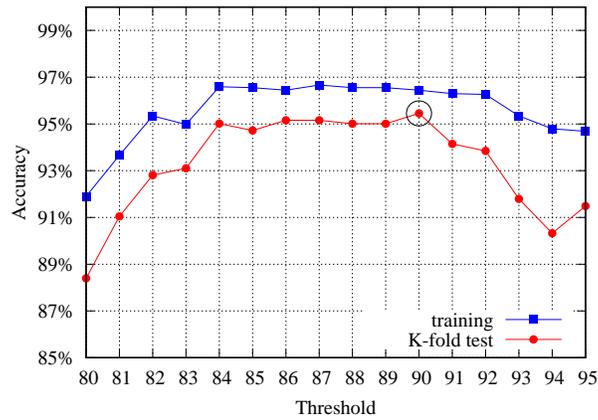}}
	\caption{The change of accuracy according to Kth iteration}
	\label{fig:fig7}
	\vspace{-0.2cm}
\end{figure}

\begin{figure}
	\centering
	\centerline{\includegraphics[width=0.5\textwidth]{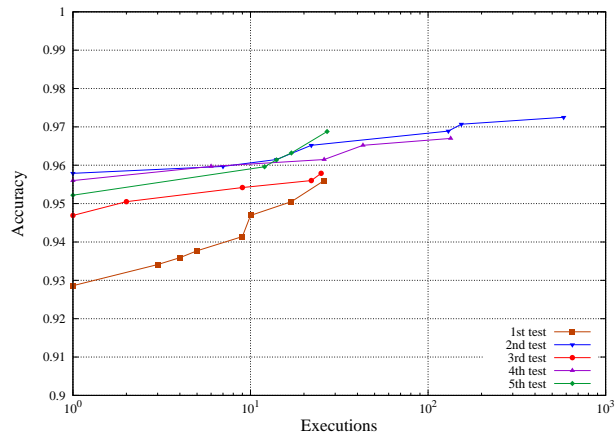}}
	\caption{K-fold cross-validation accuracy according to threshold}
	\label{fig:fig8}
	\vspace{-0.2cm}
\end{figure}

\begin{table}
	\caption{THE OPTIMAL WEIGHTS}
	\centering
	\begin{tabular}{|c|c|c|c|}
		\hline
		\begin{tabular}{@{}c@{}}\textbf{API call} \\ \textbf{sequence}\end{tabular} & \textbf{Permission} & \begin{tabular}{@{}c@{}}\textbf{Activity}\\\textbf{name}\end{tabular} & File name \\ \hline
		0.166 & 0.423 & 0.295 & 0.116 \\ \hline
	\end{tabular}
	\label{tbl:Table3}
\end{table}

We set K to 5 and the number of iterations to 1,000. As shown in Fig.~\ref{fig:fig5}, in the case an iteration of 1,000 or less, its accuracy converges to a final accuracy. Thus, setting iteration to 1,000 can use time effectively for training model. Therefore, 5,000 iterations are performed on each threshold to obtain the final accuracy. As shown in Fig.~\ref{fig:fig7}, the classification accuracy for the training set shows a slight difference for each threshold. The accuracy of the prediction is the highest when we set the optimal threshold to 90.

Therefore, the optimal threshold is set at 90. It results accurate machine learning by the training set and high prediction. Fig.~\ref{fig:fig8} shows the change in accuracy for each kth training set when the threshold is 90. The accuracy of classification is more than 95\% according to the kth training set. The prediction accuracy of the test set for kth iteration is shown in Table.\ref{tbl:Table2}. When we set the threshold to 90, Andro-Simnet have the final prediction accuracy as 95.45\%, that was the highest value. 

\begin{figure}
	\centering
	\centerline{\includegraphics[width=0.4\textwidth]{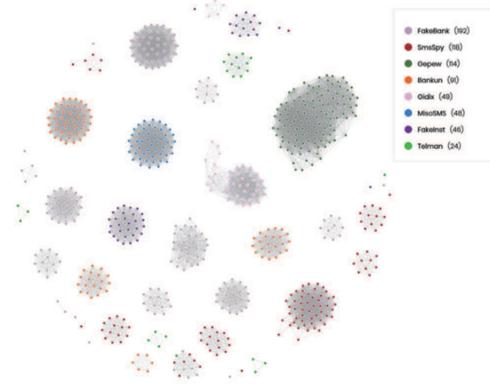}}
	\caption{The final network graph}
	\label{fig:fig9}
	\vspace{-0.2cm}
\end{figure}

\begin{figure}
	\centering
	\centerline{\includegraphics[width=0.4\textwidth]{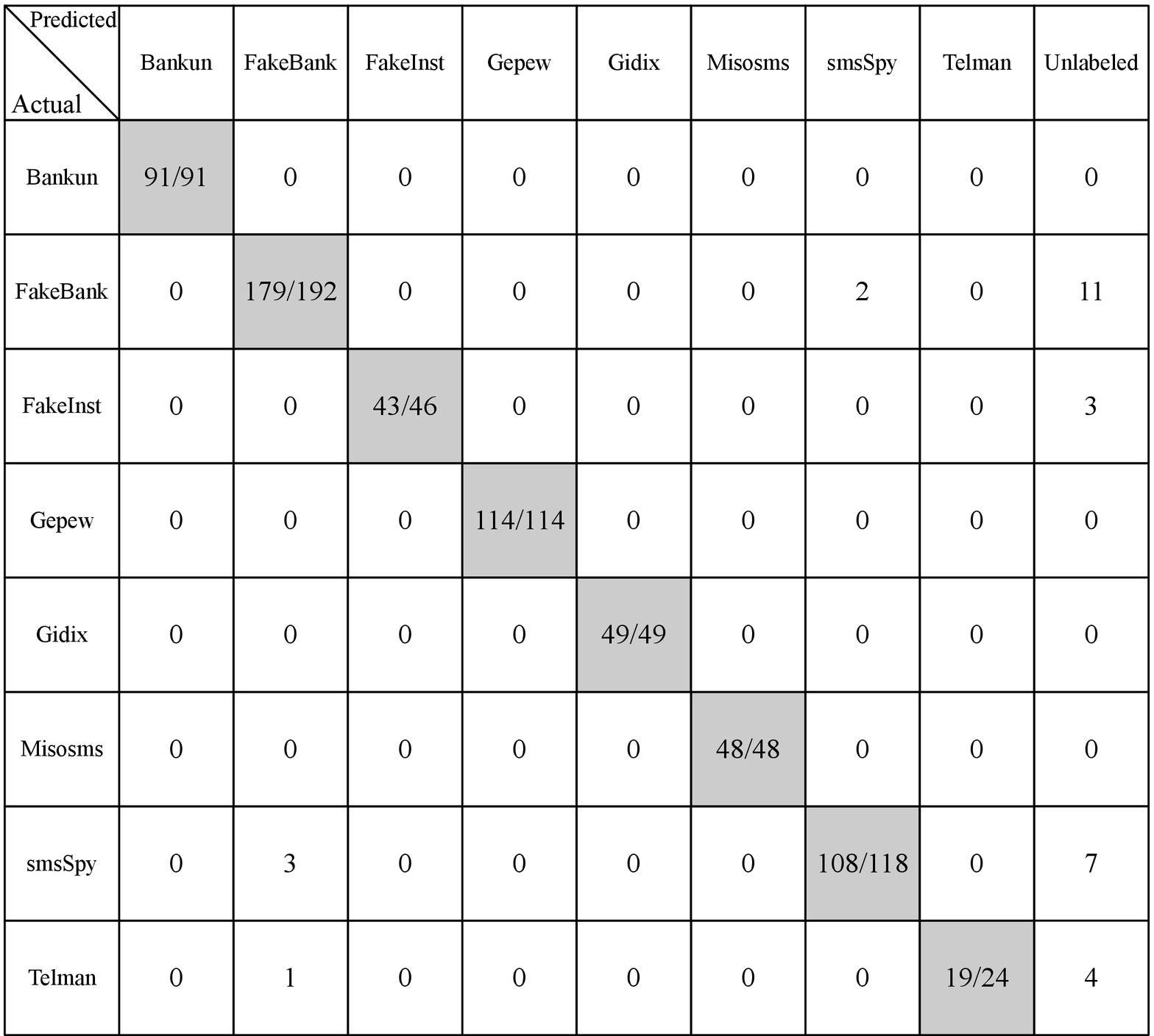}}
	\caption{Confusion matrix of K-fold cross-validation}
	\label{fig:fig10}
	\vspace{-0.2cm}
\end{figure}

Finally, we obtain the optimal weights, which are described in Table. 3. The network graph created by these weights is shown in Fig.~\ref{fig:fig9}, and the confusion matrix of K-fold cross-validation is expressed in Fig.~\ref{fig:fig10}. The process of the Andro-Simnet is shown in the video \cite{demovideo}.

\textbf{limitation} 

\begin{figure}
	\centering
	\centerline{\includegraphics[width=0.5\textwidth]{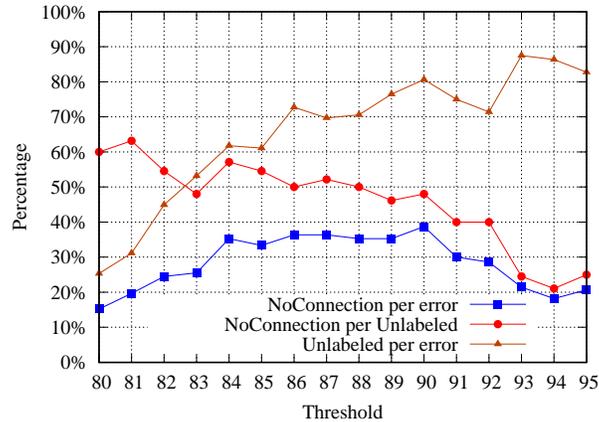}}
	\caption{Confusion matrix of K-fold cross-validation}
	\label{fig:fig11}
	\vspace{-0.2cm}
\end{figure}

The limitation of our experiment is that the prediction accuracy is not very high compared other papers. To find out the reason for low accuracy, the analysis of the experimental results was carried out. It was occurred due to 12 samples that were classified as Unlabeled in all K-fold cross-validation experiments. We diagnose a community that is composed of one node as Unlabeled. We regard these nodes as having no similarity with other nodes. 12 samples are named as NoConnection, and we analyze how they affected the prediction accuracy as shown in Fig.~\ref{fig:fig11}.
As shown in Fig.~\ref{fig:fig11}, the number of Unlabeled per errors increases as the threshold increases. The number of nodes that make up the community alone increases because the connections between nodes are severed by process of Optimizer. In most experiments, the number of Unlabeled accounts for more than 60\%. It means that the Andro-Simnet diagnoses the samples Unlabeled rather than misjudge as another family. It derives the result that NoConnection directly affects these experiments negatively. It can be seen that NoConnection occupies more than 20\% in most experiments. Also, the proportion of NoConnection occupies an average of 50\%. This shows that it causes errors in all experiments. If they are excluded from the experiment, the higher accuracy is resulted using the Andro-Simnet. To improve performance, it is helpful to obtain samples that have a connection with these 12 samples. 

\section{Conclusion}

In this paper, we use information obtained from the static analysis and dynamic analysis to extract malware’s behavioral characteristics. We can measure the similarity value between malware samples with the collected data. Using these similarity values, we do not use the general weights that are just divided equally for average, but we use optimal weights for weighted sum through the machine learning process. We think that this similarity estimation and optimal weights derivation are the key processes of clustering similar malware. Besides, the Andro-Simnet also uses community detection algorithm which can express the relationship between malware’s behavioral characteristics better. Also, the Andro-Simnet can obtain the optimal weights using social network analysis method. With the proposed machine learning algorithm, we can describe the importance value for each feature. Compared with a heuristic decision for weight value setting, Andro-Simnet can classify malware without expert knowledge. It helps to decide whether the selected feature is useful to identify associations of malware samples.

Finally, we test the Andro-Simnet with the real malware dataset to validate the performance. As a result, the Andro-Simnet can find out the most similar malware when a new sample is given. We also provide visualization results represented by a force-directed graph, which helps the security administrator and analysts to understand the results intuitively.

\section*{Acknowledgment}

This work was supported by Samsung Electronics Software R\&D Center, 2018. 

\bibliographystyle{unsrt}  
\bibliography{references}

\begin{thebibliography}{10}

\bibitem{ahmadi2016novel}
Mansour Ahmadi, Dmitry Ulyanov, Stanislav Semenov, Mikhail Trofimov, and
  Giorgio Giacinto.
\newblock Novel feature extraction, selection and fusion for effective malware
  family classification.
\newblock In {\em Proceedings of the Sixth ACM Conference on Data and
  Application Security and Privacy}, pages 183--194. ACM, 2016.

\bibitem{schmidt2009static}
A-D Schmidt, Rainer Bye, H-G Schmidt, Jan Clausen, Osman Kiraz, Kamer~A Yuksel,
  Seyit~Ahmet Camtepe, and Sahin Albayrak.
\newblock Static analysis of executables for collaborative malware detection on
  android.
\newblock In {\em 2009 IEEE International Conference on Communications}, pages
  1--5. IEEE, 2009.

\bibitem{zhang2014semantics}
Mu~Zhang, Yue Duan, Heng Yin, and Zhiruo Zhao.
\newblock Semantics-aware android malware classification using weighted
  contextual api dependency graphs.
\newblock In {\em Proceedings of the 2014 ACM SIGSAC conference on computer and
  communications security}, pages 1105--1116. ACM, 2014.

\bibitem{dataset}
Andro-simnet dataset.
\newblock \url{http://ocslab.hksecurity.net/Datasets/androsimnet}.

\bibitem{jang2016andro}
Jae-wook Jang, Hyunjae Kang, Jiyoung Woo, Aziz Mohaisen, and Huy~Kang Kim.
\newblock Andro-dumpsys: anti-malware system based on the similarity of malware
  creator and malware centric information.
\newblock {\em computers \& security}, 58:125--138, 2016.

\bibitem{ding2016android}
Yuxin Ding, Siyi Zhu, and Xiaoling Xia.
\newblock Android malware detection method based on function call graphs.
\newblock In {\em International Conference on Neural Information Processing},
  pages 70--77. Springer, 2016.

\bibitem{zhong2012malware}
Yang Zhong, Hirofumi Yamaki, and Hiroki Takakura.
\newblock A malware classification method based on similarity of function
  structure.
\newblock In {\em 2012 IEEE/IPSJ 12th International Symposium on Applications
  and the Internet}, pages 256--261. IEEE, 2012.

\bibitem{dash2016droidscribe}
Santanu~Kumar Dash, Guillermo Suarez-Tangil, Salahuddin Khan, Kimberly Tam,
  Mansour Ahmadi, Johannes Kinder, and Lorenzo Cavallaro.
\newblock Droidscribe: Classifying android malware based on runtime behavior.
\newblock In {\em 2016 IEEE Security and Privacy Workshops (SPW)}, pages
  252--261. IEEE, 2016.

\bibitem{pehlivan2014analysis}
U{\u{g}}ur Pehlivan, Nuray Baltaci, Cengiz Acart{\"u}rk, and Nazife Baykal.
\newblock The analysis of feature selection methods and classification
  algorithms in permission based android malware detection.
\newblock In {\em 2014 IEEE Symposium on Computational Intelligence in Cyber
  Security (CICS)}, pages 1--8. IEEE, 2014.

\bibitem{faruki2013androsimilar}
Parvez Faruki, Vijay Ganmoor, Vijay Laxmi, Manoj~Singh Gaur, and Ammar Bharmal.
\newblock Androsimilar: robust statistical feature signature for android
  malware detection.
\newblock In {\em Proceedings of the 6th International Conference on Security
  of Information and Networks}, pages 152--159. ACM, 2013.

\bibitem{CuckooDroid}
Cuckoodroid.
\newblock \url{http://cuckoo-droid.readthedocs.io/}.

\bibitem{damiani2004open}
Ernesto Damiani, Sabrina De~Capitani di~Vimercati, Stefano Paraboschi, and
  Pierangela Samarati.
\newblock An open digest-based technique for spam detection.
\newblock {\em ISCA PDCS}, 2004:559--564, 2004.

\bibitem{blondel2008fast}
Vincent~D Blondel, Jean-Loup Guillaume, Renaud Lambiotte, and Etienne Lefebvre.
\newblock Fast unfolding of communities in large networks.
\newblock {\em Journal of statistical mechanics: theory and experiment},
  2008(10):P10008, 2008.

\bibitem{bostock2011d3}
Michael Bostock, Vadim Ogievetsky, and Jeffrey Heer.
\newblock D$^3$ data-driven documents.
\newblock {\em IEEE transactions on visualization and computer graphics},
  17(12):2301--2309, 2011.

\bibitem{demovideo}
The demo video of andro-simnet.
\newblock \url{https://youtu.be/JmfS-ZtCbg4}.

\end{thebibliography}
\end{document}